\documentclass[twocolumn,english,notitlepage,superscriptaddress,nofootinbib]{revtex4-1}
\usepackage[T1]{fontenc}
\usepackage[latin9]{inputenc}
\usepackage{float}
\usepackage{amsmath}
\usepackage{amssymb}
\usepackage{graphicx}
\usepackage{wasysym}
 \usepackage{color}
\makeatletter

\newcommand{\affone}{Department of Physics and Astronomy, University College London, Gower Street, WC1E 6BT London, United Kingdom.}

\makeatother

\begin{document}
\title{Controlling mode orientations and  frequencies  in levitated cavity optomechanics}
\author{A. Pontin, H. Fu, J.H. Iacoponi, P.F. Barker and T.S.  Monteiro}

\affiliation{\affone}
\begin{abstract}
Cavity optomechanics offers quantum cooling, control and measurement of small mechanical oscillators. However the  optomechanical backactions disturb the oscillator motions, shifting their frequencies and generating hybridisation. This is especially relevant to levitated cavity optomechanics, where optical trapping also determines the mechanical modes and their frequencies. Here, using  a nanoparticle trapped in a tweezer in a cavity populated only by coherently scattered (CS) photons, we investigate experimentally the $S_{xy}(\omega)$ mechanical cross correlation spectra as a function of the nanoparticle position on the cavity standing wave. We show that the CS field not only rotates the mechanical modes in 
the opposite direction to the cavity dynamics, but also  opposes optical spring effects on the mechanical frequencies.
Hence we demonstrate a cancellation point, independent of most experimental parameters,  where it becomes possible to  cavity cool and control  {\em unperturbed}  mechanical modes.  The findings have implications for directional force sensing using  the CS set-ups that  permit quantum ground state cooling.
\end{abstract}
\maketitle


There is currently intense interest in optical cooling of levitated nanoparticles both with cavities as well as active feedback methods~\cite{millen2020optomechanics}.  Cooling with using optical cavities was proposed well over a decade ago, \cite{romero2010toward,barker2010cavity,chang2010cavity,pender2012optomechanical,monteiro2013dynamics}
but only in 2020 was ground state cooling achieved ~\cite{delic2020cooling}: initial experimental efforts were plagued  by technical difficulties of stable trapping in high vacuum~\cite{kiesel2013cavity,asenbaum2013cavity}. Several set-ups were investigated for cavity cooling  including tweezer-cavity traps~\cite{romero2010toward,mestres2015cooling}, electro-optical traps~\cite{millen2015cavity,fonseca2016nonlinear}, and trapping in the near field of a photonic crystal~\cite{magrini2018near}. Quantum cooling was also previously achieved with feedback cooling and quantum control~\cite{Magrini2021quantum,Novotny2021quantum}, but here we focus on levitated cavity optomechanical systems.

Recently, a 3D coherent scattering (CS) setup was introduced to levitated cavity optomechanics~\cite{delic2019cavity,windey2019cavity} using methods adapted from atomic physics~\cite{vuletic2000laser,vuletic2001three,domokos2002collective,leibrandt2009cavity,hosseini2017cavity}. In contrast to experiments that consider dispersive coupling, here the cavity is driven solely by the dipole radiation of the optically trapped silica particle. The nanoparticle is trapped at the tight focus of the optical tweezer along the $z$ axis and the tweezer laser polarisation angle and waist set the orientation and frequencies $\omega^{(0)}_x,\omega^{(0)}_y$ of the mechanical motion in the $x-y$ plane. The CS set-up yielded unprecedentedly high optomechanical coupling rates $g_x, g_y$, which subsequently enabled ground-state cooling of the motion along the cavity axis and thus opened the door to levitated cavity optomechanics at  or near~\cite{delic2020cooling,Marin2021} 
quantum regimes.

\begin{figure}[ht!]
{\includegraphics[width=3.5in]{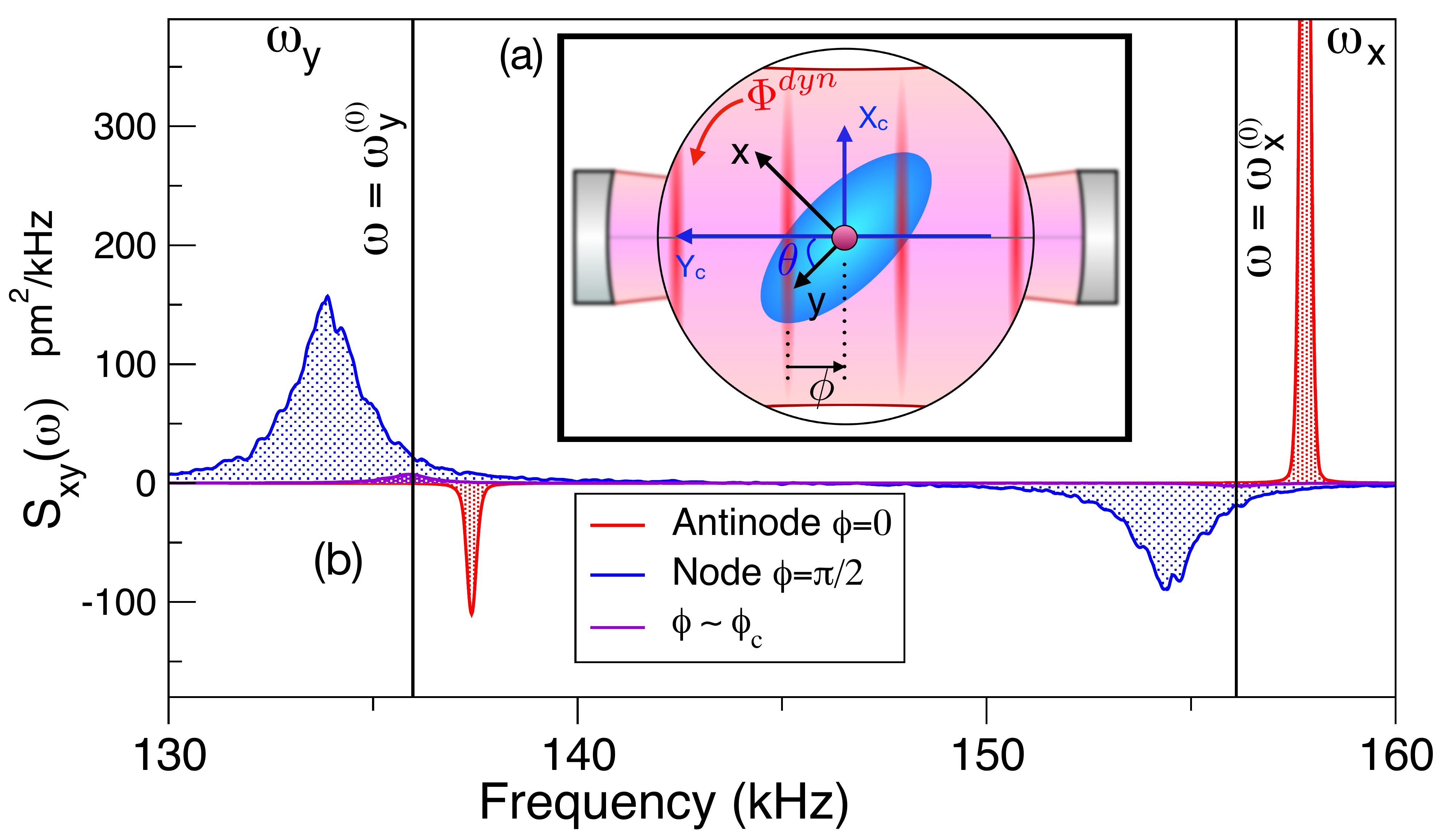}} \caption{ \textbf{(a)} Inset illustrates a nanoparticle, trapped by an optical tweezer with $x-y$ mechanical modes. However, the presence of a surrounding cavity, aimed at quantum cooling, hybridises the modes and shifts their unperturbed frequencies $\omega_{x,y}^{(0)}$. The experimental tweezer polarisation sets an initial angle $\theta$ between the $x-y$ modes and cavity axis. Then, the optomechanical cavity hybridisation dynamics adds an effective mode rotation $\Phi_{dyn}$, where $\Phi_{dyn} \sim -\theta$ leads to formation of dark/bright modes. However, away from optical nodes, the CS field  opposes this effect. (\textbf{b})  We investigate  mode orientation by measuring cross-correlation spectra $S_{xy} (\omega)$ for  $\theta \simeq \pi/4$ as the trapping position is swept from node (blue, $\phi=\pi/2$) to near the antinode (red, $\phi=0$) of the cavity standing wave. The $x-y$ motions are always anticorrelated (peaks of opposite sign), but $S_{xy}$ flips sign at $\phi=\phi_c$ (purple line, $S_{xy} (\omega)\sim 0$) . For this value, the CS field cancels $\Phi_{dyn}$: the mechanical modes are locked at their unperturbed orientations {\em and} unperturbed frequencies for arbitrary power and  $\theta$, but can still be strongly cooled. The results have implications for directional force sensing and, in strong coupling regimes, the suppression of dark/bright modes.  }
\label{Fig1}
\end{figure}

\begin{figure*}[ht]
{\includegraphics[width=6.5in]{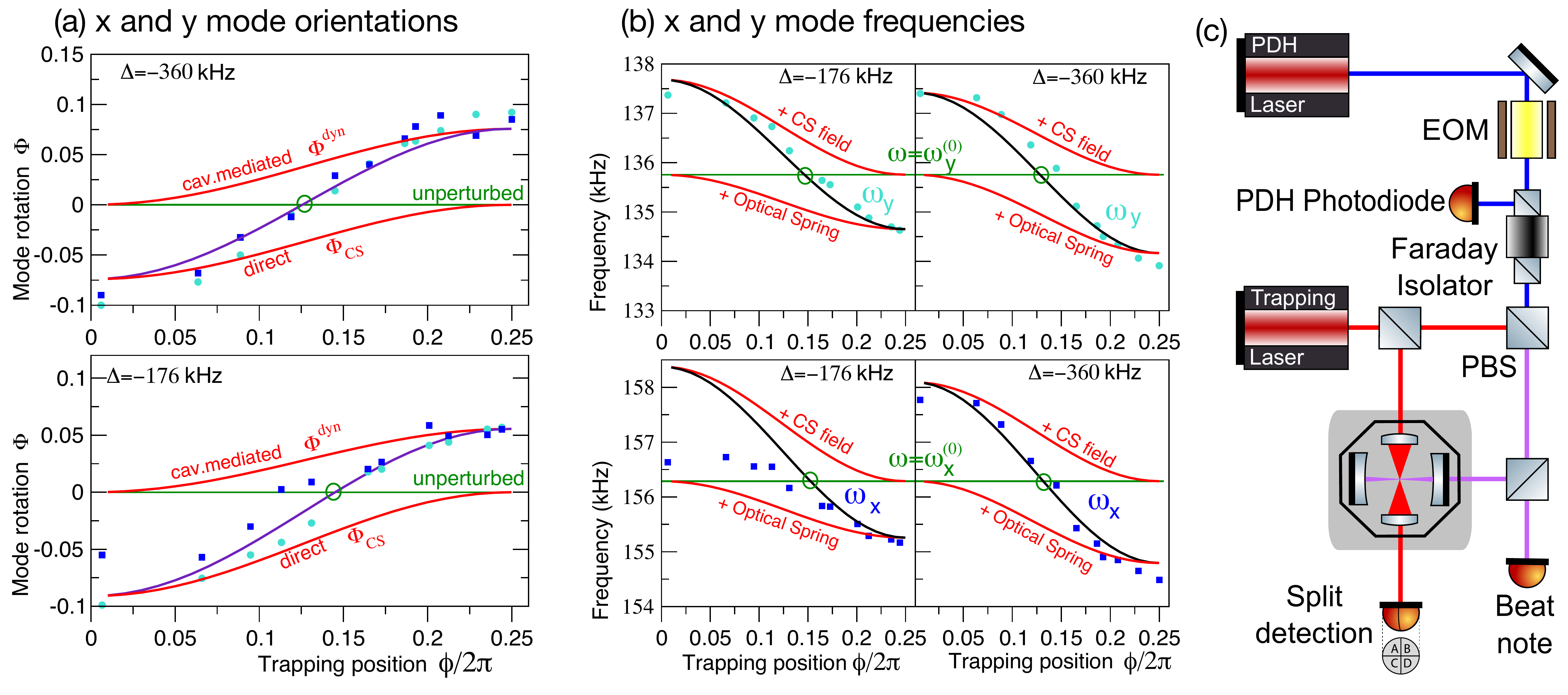}} \caption{ \textbf{(a)} Measured orientation of the mechanical modes in $x-y$ plane as a function of trapping position.  The mode rotation  angles, $\Phi$, are  obtained from  experimental cross  correlation spectra, $S_{xy}(\omega)$:  squares (blue) correspond to the $x$ peak, circles (cyan) to the $y$ peak.   $\Phi$  combines a positive sign contribution, 
$\Phi_{dyn}$,  from the optomechanical dynamics arising from optical  backaction, and $\Phi_{CS}$, a negative sign contribution from the static CS field. $\Phi_{dyn}$ is dominant at the node ($\phi=\pi/2$) while  $\Phi_{CS}$ dominates at the antinode ($\phi=0$). Where they cancel (green circle, $\phi=\phi_c$ ) the rotation becomes zero. The measured values of $\Phi$ are compared  with $\Phi_{CS}+\Phi_{dyn}$ calculated from theory (violet lines). \textbf{(b)} Shows that the frequency behavior mirrors the mode rotation behavior: the optomechanical optical spring effect  softens the mechanical frequencies for red detuning, while in contrast CS field  stiffens the frequencies.
At $\phi\simeq \phi_c$ the two contributions cancel, leaving only the unperturbed mechanical frequency set by the tweezer trap.  Red lines denote theoretical values, calculated from linearised optomechanical equations of motion. Black line shows their combined effect. $\theta \simeq \pi/4$ for all plots.
\textbf{(c)} Experimental set-up, described in the text.}
\label{Fig2}
\end{figure*}

Theoretical analysis  \cite{MTTM2020,MTTM2021} has shown that $x-y$ hybridisation results from  both cavity mediated, back action terms and a direct coupling of the form $g_{xy} \hat{x}\hat{y}$. There can be three-way hybridisation  of the normal modes in the $x-y$ plane as well as hybridisation with optical modes, especially in strong-coupling regimes. The interesting case of $g_{x}\simeq g_{y}$ which is obtained experimentally for tweezer polarisation angle $\theta=\pi/4$, in the strong-coupling regime produces a mode rotation angle of $\Phi \sim \pi/4$ and leads to the formation of dark and bright modes~\cite{Harris2014,MTTM2021,Marin2021}.

The set-up is illustrated in {\bf Fig.\ref{Fig1}} :  we investigate the  behavior of mechanical modes  in the $x-y$ plane experimentally by measuring cross-correlation $S_{xy}(\omega)$ spectra. We show that  rotation of the modes by an angle  $\Phi_{dyn}$ due to a dynamical optical backaction term is, away from the node, opposed by a rotation  $\Phi_{CS }$  due to the  CS potential. In addition, we find here for the first time, that the experimental mechanical frequency shifts directly mirror this behaviour. At the node, the optical spring effect due to the cavity dynamics softens the mechanical frequencies; away from the node, the CS potential stiffens the mechanical frequency. The two effects cancel at the same point as the cancellation of the mode rotation angle. At this point the mechanical $x,y$ frequencies both return to their unperturbed values and the modes are locked to their unperturbed orientation, for arbitrary input power and polarisation. Yet there is still strong optomechanical cavity cooling.  Trapping away from the node can be less advantageous:  cavity photon occupancies are higher  at the antinode relative the node ~\cite{delic2019cavity}.  However,  as the detuning tends to resonance, the cancellation point will move  closer to the node, mitigating this disadvantage.

{\em Experimental set-up:} A schematic overview of the experiment is shown in {\bf Fig.\ref{Fig2}(c)}. We use two Nd:YAG lasers at a wavelength $\lambda=1064$\,nm. A weak field from the first is used to implement a Pound-Drever-Hall (PDH) scheme to lock it to a high finesse Fabry-Perot cavity. The second laser illuminates an optical tweezer composed of a single aspheric lens of nominal numerical aperture $NA=0.77$ and a symmetric condenser lens. The tweezer assembly is monolithic, is mounted on a $3$-axes translational stage and includes an aspheric collection lens of $NA=0.3$, oriented at $90^{\circ}$ from the propagation direction, which is exploited for imaging purposes. The tweezer trapping region is positioned at the center of the optical cavity which has a length $L_{cav}=12.23\pm0.02$\,mm, a finesse $\mathcal{F}\simeq31000$  (linewidth $\kappa/2\pi=396\pm2$\,kHz, input rate $\kappa_{in}/2\pi=162\pm2$\,kHz) and a waist of $61\,\mu$m. The two lasers are phase offset locked by monitoring their beat note and their frequency separation is set to one free spectral range (FSR=$c/2L_{cav}= 12.26\pm0.02$\,GHz). The PDH beam is locked at resonance and the detuning of the trapping beam can be finely controlled. Importantly, the PDH beam is orthogonally polarized with respect to the tweezer beam and interacts with the particle only dispersively. This, in combination with the low power makes the role of the PDH beam in the dynamics negligible. The motion of the particle in the tweezer polarization plane is monitored by distributing the tweezer light collected by the condenser lens to two balanced detectors. Each detector is balanced using D-shaped mirrors oriented parallel and perpendicular to the beam polarization thus sensitive to $x$ and $y$ respectively. The two detectors can measure independently the $x$ and $y$ motion with a rejection ratio potentially exceeding $-30$\,dB, however, any imperfections in the orientations of the D-mirror can result in a small mixing of $x$ and $y$ in the detected signals.

{\em Physical model}: In a coherent scattering (CS) approach, the optical cavity is not externally driven but it is populated exclusively by light scattered by the nanoparticle. The corresponding Hamiltonian results from the coherent interference between the electric fields of the tweezer and cavity  $\hat{H}=-\frac{\alpha}{2}\vert\mathbf{\hat{E}}_{\text{cav}}+\mathbf{\hat{E}}{}_{\text{tw}}\vert^{2},$ where  $\alpha$ is the polarizability of the nanosphere.  The interference term $\propto (\mathbf{\hat{E}}_{\text{cav}}^{\dagger}\mathbf{\mathbf{\hat{E}}}_{\text{tw}}+\mathbf{\hat{E}}_{\text{cav}}\mathbf{\mathbf{\hat{E}}}_{\text{tw}}^{\dagger})$ gives rise to an effective
 CS potential:
 \begin{equation}
\hat{V}_{\text{CS}}/\hbar = -E_{d}\text{cos}(\phi+k \hat{Y}_c) e^{-({\hat{x}^{2}/w_{x}^{2} +\hat{y}^{2}/w_{y}^{2} })} \left[\hat{a}+\hat{a}^{\dagger}\right].
\label{eq:VCS}
\end{equation}

In the above, we omitted the $\hat{z}$ dependence. The CS dynamics decouples into a 2+1 dynamics with $x-y$ motion close in frequency and prone to hybridise; and a (typically) off-resonant $z$ motion that is largely decoupled. We note that the $z$ motion can be strongly cooled using feedback cooling~\cite{Magrini2021quantum,Novotny2021quantum} but, in most experimental implementations, $z$ is anyway moderately cooled due to a small unavoidable tilt relative to the cavity $z$ axis. This effect and the full $z$ dynamics is fully taken into account in numerical simulations, however, for brevity, it is not included in the present discussion.

In Eq.\,\ref{eq:VCS}, $Y_c$ represents the cavity axis. Relating the cavity coordinates to the tweezer frame is a simple rotation of the coordinate frames $[X_c \ Y_c]^T= R_z(\theta) [x \ y]$, where $R_z(\theta)$ is the 2D rotation matrix. $\phi$ is the displacement of the trap from an antinode,  $w_y\simeq1.068 \mu$m and $w_x\simeq0.928 \mu$m are the tweezer waists.

Linearised equations of motion are obtained using Eq.\ref{eq:VCS}; operators are expanded about equilibrium values, thus the optical field operators $\hat{a}\to \overline{\alpha} + \hat{a}(t)$ are expanded about the mean field, where $n_p=| \overline{\alpha}|^2$ is the mean photon occupancy of the cavity. This linearisation yields also optomechanical coupling terms $g_x\hat{x}(\hat{a}+ \hat{a}^{\dagger})$ and $g_y \hat{y}(\hat{a}+\hat{a}^{\dagger})$. In~\cite{MTTM2020,MTTM2021} it was shown further that the CS potential also yields direct coupling terms $g_{xy} \hat{x}\hat{y}$ that are of similar order to the usual linearised terms, thus do not vanish as the quantum occupancies are approached.

In ~\cite{MTTM2020,MTTM2021} it was shown that $x-y$ hybridisation requires a correction to the 1D unhybridised 
solutions, hence  $\hat{x}(\omega) \simeq \hat{x}^{1D}(\omega)+\mathcal{R}_{xy}(\omega)\hat{y}(\omega)$ and $\hat{y}(\omega) \simeq \hat{y}^{1D}(\omega)+\mathcal{R}_{yx}(\omega)\hat{x}(\omega)$,
where $\mathcal{R}_{xy},\mathcal{R}_{yx}$ are hybridisation functions that are small in the weak coupling regimes of the present experiments. In that case if  $\mathcal{R}_{xy}\simeq -\mathcal{R}_{yx}$ the above linear $x-y$ hybridisation relation is already suggestive of a simple frame rotation.

One can show (see Appendix for details) that the resulting mechanical correlation spectra:
\begin{equation}
S_{xy} (\omega) \approx  \text{Re} ({\mathcal{R}_{yx}}(\omega)) S_{xx}(\omega)  +\text{Re}({\mathcal{R}_{xy}}(\omega)) S_{yy} (\omega)
\label{Corr1}
\end{equation}
depend on the real parts of the hybridisation functions. For $\omega \sim \omega_{x,y}$, one can show  $\mathcal{R}_{xy}\simeq -\mathcal{R}_{yx}\equiv G(\omega)/(\omega_x-\omega_y)$  where $G(\omega)= \left[i\eta_{c}(\omega)g_{x}g_{y}+g_{xy}\right]$
and $\eta_{c}= \chi(\omega)-\chi^*(-\omega)$, $\chi(\omega)=1/(-i(\omega+\Delta)+\kappa/2)$  is the cavity susceptibility function.
 Hence we can write:

\begin{equation}
S_{xy} (\omega) \simeq  \Phi  \  [  S_{yy}(\omega) - S_{xx}(\omega) ].
\label{modeangle}
\end{equation}

The above expression is generic to a 2D optomechanical system with non-zero $g_x,g_y,g_{xy}$ in many typical regimes. However, the CS version has additional and unexpected features. The angle $\Phi= \Phi_{dyn} +\Phi_{CS}$ can be decomposed into two separate contributions: (i) a cavity mediated term $\Phi_{dyn} \equiv \text{Re} ({i\eta_{c}g_{x}g_{y}})/(\omega_x-\omega_y)$  and (ii) a direct contribution $\Phi_{CS} \equiv g_{xy}/(\omega_x-\omega_y)$ arising from the static CS potential. It was found in~\cite{MTTM2020} that the optomechanical couplings  are $g_x \simeq -E_d\ k  \sin \theta \sin \phi X_{zpf}$,
$g_y \simeq  -E_d\ k  \cos \theta \sin \phi Y_{zpf}$ and

\begin{equation}
g_{xy}\simeq  g_{x}g_{y} 2\Delta\cot^2{\phi}/(\kappa^2/4 +\Delta^2)
\end{equation}
thus the combined dynamical and CS rotation becomes:

\begin{equation}
 \Phi=\Phi_{dyn}+\Phi_{CS}= \frac{g_xg_y}{(\omega_x-\omega_y)}\left[\text{Re}({i\eta_{c}})+ \frac{2\Delta\cot^2{\phi}}{(\kappa^2/4 +\Delta^2)}\right].
 \label{eq:Phi}
\end{equation}
The two terms are of oppposite sign, hence their effect is to rotate the modes in opposing directions.
 $X_{zpf}=\sqrt{\hbar/(2m\omega_x)}$, $Y_{zpf}=\sqrt{\hbar/(2m\omega_y)}$.

The experimental PSDs $S_{xx}, S_{yy}$ and correlation spectra at different trap positions are acquired at a constant pressure of $3\times10^{-3}$\,mbar with each time trace covering an observation time of $10$\,s. From these we obtain $\Phi$ as a function of the trapping positions $\phi$.

The measured rotation in the mode orientations are shown in {\bf Fig.\ref{Fig2}(a)} for two separate detunings which represent the two interesting limit cases.  To model the experiments, for all results, we employed nanosphere radius $R=60.1$\,nm,  input power to the tweezer  $P_{tw}= 0.485$\,W  and  $\theta=49^\circ$. \\

{\em Behavior of the mechanical frequencies:}  An interesting and unexpected observation is that the behavior of the frequencies mirrors the mode rotations; at the $\phi \simeq \phi_c $ position, they return to their unperturbed values. This is shown in {\bf Fig.\ref{Fig2}(b)}, for two values of the optical detuning. Here, the experimental values are obtained by fitting the PSDs.

The unperturbed mechanical frequencies of this levitated optomechanical system are set by the tweezer trap:
\begin{equation}
 (\omega_{(x,y)}^{(0)})^2 = \frac{\alpha \epsilon^2_{tw}}{mw_{x,y}^2}
 \label{eq:twtrap}
\end{equation}
where $\epsilon^2_{tw}=4P_{tw}/(w_xw_y \pi c\epsilon_0)$ is related to the input power from the tweezer.
In the presence of the cavity the coupling to the optical mode dynamics introduces an optical spring `softening' (for red-detuned light) that is generic to all cavity optomechanical set-ups. In strong coupling regimes this can be a very large shift. Neglecting a correction for 2D $x-y$ coupling \cite{MTTM2020,MTTM2021}, the optical spring shift $ (\delta\omega^{(j)}_{OS})^2= \text{Re}({-i 2g^2_j \omega_{j}^{(0)}  \eta_{c}}) $ with $j=x,y$.\\

However, for the coherent scattering set-up, there is a countervailing potential, obtained by linearising Eq.\ref{eq:VCS}, that `stiffens' the mechanical frequencies.  It is a static contribution, dependent on the mean photon occupancy of the cavity, thus can be considered an effect of co-trapping by the CS field. It takes the form $(\delta\omega^{(j)}_{CS})^2 \simeq  -(E_d k^2 \sin^2 \theta/m)  2\Delta \cos^2\phi [(\kappa/2)^2+ \Delta^2]^{-1}$.

Hence the corrected frequencies combine the two contributions: $(\omega_j)^2=(\omega_j^{(0)})^2+(\delta\omega^{(j)}_{CS})^2+ (\delta\omega^{(j)}_{OS})^2$  and may be written:

\begin{equation}
 (\omega_j)^2\simeq (\omega_j^{(0)})^2-2g_j^2\omega_j^{(0)}\left[\text{Re}({i\eta_{c}})+ \frac{2\Delta\cot^2{\phi}}{(\kappa^2/4 +\Delta^2)}\right]
 \label{eq:freq}
\end{equation}

\begin{figure}[t]
{\includegraphics[width=3.3in]{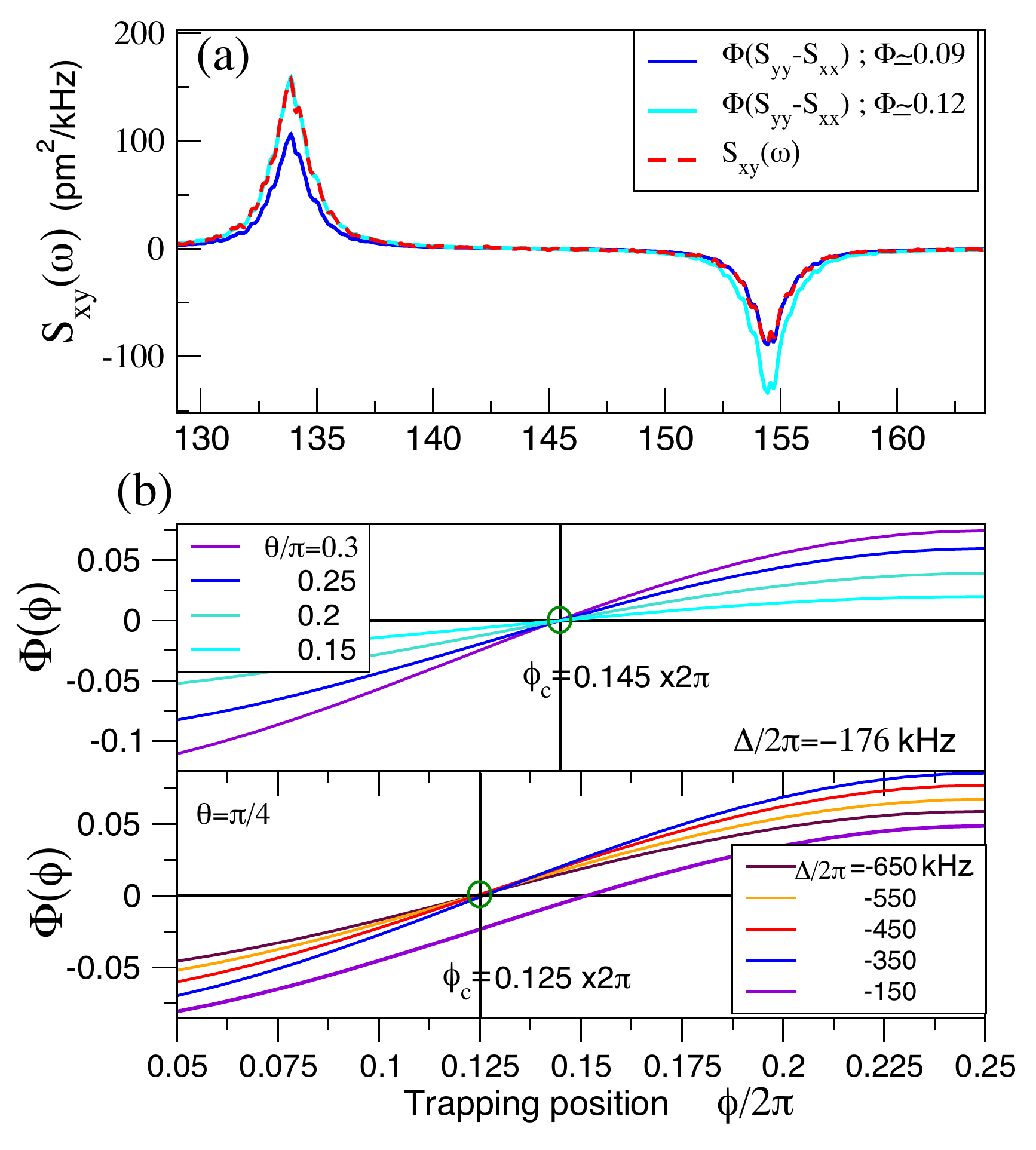}} \caption{  \textbf{(a)} To measure mode rotation, we compare cross correlation spectra $S_{xy} (\omega)$  with the rescaled difference between PSDs   $S_{xy} (\omega) \simeq  \Phi (\phi) \  [  S_{yy}(\omega) - S_{xx}(\omega) ]$ to extract $ \Phi (\phi)$. Plots show experimental $S_{xy}$ at a node, $\Delta/2\pi=-360$ kHz and rescaled, measured, $S_{yy}-S_{xx}$. The rescaling gives excellent agreement, but there is a bias $ \Phi_\beta \approx 0.03$ arising from an imperfection in the orientation of the D-mirror ($\sim2^{\circ}$) for the $y$ detector. This yields a systematic shift between the scaling of all points obtained from the $y$ peak (cyan) and the $x$ peak (dark blue). In Fig.\ref{Fig2} all $y$ (cyan) data points are  shifted by a constant,  $\Phi (\phi)\to \Phi (\phi)-\Phi_\beta$.  Upper red curve shows $x-y$ hybridisation mediated by the cavity mode.  \textbf{(b)} Illustrates the `locking' of the mode orientation at $\phi=\phi_c$.
(i) (upper panel): for $\Delta/2\pi=-176$ kHz, nearing resonance at $ \omega_{x,y}/(2\pi) \sim 150$ kHz, $\phi_c/2\pi = 0.145$. At this point, for arbitrary tweezer polarisation $\theta$ or input power, the modes remain at the unperturbed orientations.  (ii) (lower panel) for large detuning  $-\Delta/2\pi >350$ kHz the locking point is $\phi_c/2\pi=0.125$.
For lower detunings, $\phi_c$ moves towards the node and at $\Delta/2\pi=-150$ kHz, $\phi_c/2\pi\simeq 0.15$.  }
\label{Fig3}
\end{figure}

It is clear from a comparison with Eq.\ref{eq:Phi} that the multiplicative terms in square brackets are equivalent; thus if they cancel for the mode orientations, they cancel for the frequency contributions.  A more refined analysis, with consideration of weaker corrections and higher order terms will show that the $\phi=\phi_c$ point is not identical for the frequencies and orientations, but it is quite close.

In {\bf Fig.\ref{Fig3}}(a), the procedure for measuring the mode angles $\Phi$ as a function of $\phi$ is illustrated.  A small
imperfection in the orientation of the D-mirror, of the order of $2^{\circ}$, for the detection of $y$ introduces a nonvanishing cross correlation with $x$ even in absence of the cavity interaction. This translates into a constant bias in the evaluation of $\Phi$, which can be easily removed.

{\em Behavior of $\phi_c$}: the position of the cancellation point is  shown in {\bf Fig.\ref{Fig3}}(b). Clearly, from Eq.\ref{eq:Phi} and Eq.\ref{eq:freq}, there is no dependence of $\phi_c$ on the experimental polarisation or input power. The upper panel shows this behavior for $\Delta/2\pi=-176$ kHz. The rate of rotation of the mode orientation varies considerably, but the cancellation point remains fixed at $\phi_c/2\pi=0.145$. The lower panel shows that for large detuning, $\phi_c$  is independent of $\Delta$. As shown in~\cite{MTTM2020},  at large detuning, $\text{Re}({i\eta_{c}}) \to \frac{-2\Delta}{(\kappa^2/4 +\Delta^2)}$ is real, we obtain $1-2 \cot^2 \phi_c=0$, implying $\phi_c/2\pi=0.125$ for all $\Delta$. However,  at  resonance   ($-\Delta \sim \omega_{x,y}$),  $\phi_c$ moves towards the node. For this particular cavity, $\phi_c/2\pi \simeq 0.15$ near resonance. For lower $\kappa$ and below resonance,  one can approach $\phi_c/2\pi=0.2$,
where the cavity photon occupancy is only $\sim  4-5$  times larger than at the node. But Eq.\ref{eq:Phi} assumes weak coupling;
a more detailed analysis is required to obtain the $\Phi \simeq 0$ point in regimes with significant mechanical-optical hybridisation.

{\em Conclusions}  CS set-ups approaches have had a disruptive effect on levitated cavity optomechanics for two key reasons: (i) they offer  optomechanical coupling strengths  an order of magnitude larger relative to earlier dispersive approaches and (ii) the possibility of trapping at a cavity node, not possible in dispersive set-ups, offers an elegant solution to the detrimental impact of technical noise, including classical laser frequency noise, because of the much lower cavity photon occupancy.
Hence most CS experiments trap  at a cavity node. However, a theoretical study \cite{MTTM2020} indicated trapping away from the node (for $-\Delta \gg \omega_{x,y}$, might offer a regime where the CS potential cancels  the dynamics-induced hybridisation of the mechanical modes. 

Here this effect is demonstrated and investigated experimentally. In addition, it was also found, for the first time, that the frequency shifts precisely mirror this effect and that the cancellation point, for lower detunings,  moves closer to the node.

 Future studies 
 will combine technical improvements to the frequency noise with operation closer to (but not exactly at) the node and will
 thus be able to maximise the potential of this regime.
This has important potential consequences for sensing applications. For example, the directional sensitivity of levitated nanoparticles aids  the search for dark matter candidates since rejection of background events can be enhanced by the knowledge of the direction of the incoming dark matter candidate~\cite{dark_matter_1,dark_matter_2,dark_matter_3}.
The same process can impede the formation of bright/dark modes  in the strong coupling regime allowing efficient 2D quantum cooling.

{\em Acknowledgements} The authors would like to acknowledge helpful discussions with J.~Gosling and M.~Toro\v{s}. The authors acknowledge funding from the Engineering and Physical Sciences Research Council (EPSRC) Grant No. EP/N031105/1. H.F. and J.H.I. acknowledge
EPSRC studentship funding via grant number EP/L015242/1 (H.F.)  and EP/R513143/1 (J.H.I.).
\bibliographystyle{unsrt}
\bibliography{2Dbiblio}

\begin{Large}
\begin{center}
APPENDIX
\end{center}
\end{Large}

\section{Mechanical frequencies}

The combined tweezer-cavity Hamiltonian takes the form:
\begin{equation}
\hat{H}=-\frac{\alpha}{2}\vert\mathbf{\hat{E}}_{\text{cav}}+\mathbf{\hat{E}}{}_{\text{tw}}\vert^{2},\label{eq:Hch}
\end{equation}
where $\mathbf{\hat{E}}_{\text{cav}}$ ($\mathbf{\hat{E}}_{\text{tw}}$)
denotes the cavity (tweezer) field, $\alpha=3\epsilon_{0}V_{s}\frac{\epsilon_{R}-1}{\epsilon_{R}+2}$
is the polarizability of the nanosphere, $V_{s}$ is the volume of
the nanosphere, $\epsilon_{0}$ is the permittivity of free space,
and $\epsilon_{R}$ is the relative dielectric permittivity.

We assume a coherent Gaussian tweezer field and replace the modes
with c-numbers to find:

\begin{equation}
\mathbf{\mathbf{\hat{E}}}_{\text{tw}}=\frac{\epsilon_{tw}}{2}\frac{1}{\sqrt{1+(\frac{z}{z_{R}})^{2}}}e^{-\frac{\hat{x}^{2}}{w_{x}^{2}}}e^{-\frac{\hat{y}^{2}}{w_{y}^{2}}}e^{ik\hat{z}+i\Phi(\hat{z})}e^{-i\omega_{\text{tw}}t}\mathbf{e}_{y}+\text{cc}\label{eq:Etw}
\end{equation}
where $\Phi(z)=-\arctan\frac{z}{z_{R}}$ is the Gouy phase, $z_{R}=\frac{\pi w_{x}w_{y}}{\lambda}$
is the Rayleigh range, $w_{x}$ ($w_{y}$) are the beam waist along
the $x$ ($y$) axis, $\epsilon_{tw}=\sqrt{\frac{4P_{\text{tw}}}{w_{x}w_{y}\pi\epsilon_{0}c}}$
is the amplitude of the electric field, $c$ is the speed of light,
$P_{\text{tw}}$ is the laser power, $\omega_{\text{tw}}$ is the
tweezer angular frequency, $t$ is the time, and $\hat{\boldsymbol{r}}=(\hat{x},\hat{y},\hat{z})$
is the position of the nanoparticle. $\mathbf{e}_{j}$ are the unit
vectors: $\mathbf{e}_{z}$ is aligned with the symmetry axis of the
tweezer field and $\mathbf{e}_{y}$ is aligned with the polarization
of the tweezer field.

The cavity field is given by:

\begin{equation}
\mathbf{\hat{E}}_{\text{cav}}=\epsilon_{c}\text{cos}(k(Y_{0}^{\text{(c)}}+\hat{Y}^{\text{(\text{c)}}}))\mathbf{e}_{x}^{c}\left[\hat{a}+\hat{a}^{\dagger}\right],\label{eq:Ecav}
\end{equation}
where $\epsilon_{c}=\sqrt{\frac{\hbar\omega_{c}}{2\epsilon_{0}V_{c}}}$
is the amplitude at the center of the cavity, $V_{c}$ is the cavity
volume, $\omega_{c}$ is the cavity frequency, $\hat{a}$ ($\hat{a}^{\dagger}$
)is the annihilation (creation) operator, $Y_{0}^{\text{(c)}}$ is
an offset of the cavity coordinate system (centered at a cavity antinode)
with respect to the tweezer coordinate system.

The cavity $X_{c}$-$Y_{c}$
plane is rotated by an angle $\theta$ with respect to the tweezer
$x$-$y$ plane. For $\theta=0$, the tweezer polarization ($y$-axis) becomes
aligned with the cavity symmetry axis ($Y_{c}$-axis). In particular,
we have $\hat{Y}^{\text{(\text{c)}}}=\text{sin}(\theta)\hat{x}+\text{cos}(\theta)\hat{y}$
while $\hat{X}^{\text{(\text{c)}}}=\text{cos}(\theta)\hat{x}-\text{sin}(\theta)\hat{y}$.

We expand the Hamiltonian in Eq.~(\ref{eq:Hch}), exploiting Eqs.~(\ref{eq:Etw}) and
(\ref{eq:Ecav}) to obtain three terms:

\begin{equation}
\hat{H}=-\frac{\alpha}{2}\vert\mathbf{\mathbf{\hat{E}}}_{\text{tw}}\vert^{2}-\frac{\alpha}{2}\vert\mathbf{\hat{E}}_{\text{cav}}\vert^{2}-\frac{\alpha\text{sin}(\theta)}{2}(\mathbf{\hat{E}}_{\text{cav}}^{\dagger}\mathbf{\mathbf{\hat{E}}}_{\text{tw}}+\mathbf{\hat{E}}_{\text{cav}}\mathbf{\mathbf{\hat{E}}}_{\text{tw}}^{\dagger}),\label{eq:Hch2}
\end{equation}
where the terms on the right hand-side represent the tweezer trapping potential, the cavity
intensity field, and the tweezer-cavity interaction term (from left to right).
The first (tweezer field) term dominates the trapping and primarily
sets the three mechanical frequencies $\omega_{x}$, $\omega_{y}$,
and $\omega_{z}$. In the discussion below we focus only on the $x,y$ modes.

The second term represents  trapping by the cavity intensity potential.
In earlier experiments in levitated cavity optomechanics with no tweezer trap, this field determined the mechanical
frequencies. However, achieving reasonable frequencies required very high photon occupancies
$n_p \sim 10^{10}$.  In CS setups this term provides a negligible correction to the frequencies but is included in the numerics  for  precision.

The
third term, which we will denote as $\hat{V}_{\text{CS}}$, is the
most interesting,  as this is the coherent scattering potential that has had a transformative effect
in the levitated cavity optomechanics field:

\begin{eqnarray}
\frac{\hat{V}_{\text{CS}}}{\hbar}&=&-E_{d}\text{cos}(\phi+k(\hat{x}\sin\theta+\hat{y}\cos\theta))e^{-\frac{\hat{x}^{2}}{w_{x}^{2}}}e^{-\frac{\hat{y}^{2}}{w_{y}^{2}}}\hat{A}; \nonumber \\
&\textrm{where}& \ \hat{A}=\left[\hat{a}e^{-i(k\hat{z}+\Phi(\hat{z}))}+\hat{a}^{\dagger}e^{+i(k\hat{z}+\Phi(\hat{z}))}\right]
\label{eq:VCS2}
\end{eqnarray}

\begin{equation}
\textrm{and} \  E_{d}=\frac{\alpha\epsilon_{c}\epsilon_{tw}\sin\theta}{2\hbar},
\end{equation}
while $\phi=kY_{0}^{\text{(c)}}$ represents the effect of the shift between
the origin of the tweezer field and the cavity standing wave.
Linearising $\hat{V}_{\text{CS}}$ about  equilibrium displacements  yields 
 the optomechanical coupling strengths:
\begin{eqnarray}
g_x &\simeq & -E_d\ k  \sin \theta \sin \phi X_{zpf} \nonumber\\
g_y &\simeq & -E_d\ k  \cos \theta \sin \phi Y_{zpf},
\end{eqnarray}
 $X_{zpf}=\sqrt{\hbar/(2m\omega_x)}$, $Y_{zpf}=\sqrt{\hbar/(2m\omega_y)}$.
 
In addition, as a feature of  the CS set up, the linearisation yields also direct $x-y$ couplings  $g_{xy}\simeq-g_{x}g_{y}\frac{2\text{Re}(\bar{\alpha})\cos{\phi}}{E_{d}\sin^{2}{\phi}}$ related to the mean photon occupancy $n_p= |\bar{\alpha}|^2$  in the cavity. 
They are negligible at the nodes ($\phi=\pi/2$), but become stronger as $\phi \to 0$.

\subsection{tweezer trap frequencies}

The mechanical $x,y$ frequencies are set mainly by the tweezer trap $-\frac{\alpha}{2}\vert\mathbf{\mathbf{\hat{E}}}_{\text{tw}}\vert^{2}$. Linearising leads to

\begin{equation}
 (\omega^{(0)}_{x,y})^2 = \frac{\alpha \epsilon^2_{tw}}{mw_{x,y}^2}
 \label{eq:twtrap}
\end{equation}

\subsection{co-trapping}
The linearisation of ${\hat{V}_{\text{CS}}}$ also yields corrections to the tweezer trap frequencies: 
the zero-th order frequency must be corrected by co-trapping by the coherent scattering potential in
Eq.(\ref{eq:VCS2}). These stiffen the tweezer frequencies by corrections of the form:

\begin{eqnarray}
(\delta \omega_{x})^2 &\simeq & \frac{E_d\hbar}{m} 2 \alpha_R \cos \phi [k^2  \sin^2\theta +\frac{2}{w_x^2}] \nonumber\\
(\delta \omega_{y})^2 &\simeq & \frac{E_d\hbar}{m} 2 \alpha_R \cos \phi [k^2  \cos^2\theta +\frac{2}{w_y^2}]
\label{CSshift}
\end{eqnarray}

For the $x,y$ frequencies, the stiffening depends on the mean cavity field:
\begin{equation}
\overline{\alpha}=\alpha_R+i\alpha_I = \frac{-iE_d\cos \phi}{i\Delta-\kappa/2}
\end{equation}
so $\alpha_R=\frac{-\Delta E_d \cos \phi }{\kappa^2/4 +\Delta^2}$.

We note that the second term in the square brackets in Eq.\ref{CSshift} is small, so
$(\delta \omega_{x})^2 \simeq  \frac{E_d\hbar}{m} 2 \alpha_R \cos \phi k^2  \sin^2\theta $ represents already a good
approximation to the frequency correction.

\subsection{optical spring}

A well-studied effect in optomechanics is the so called `optical spring' shift of the mechanical frequencies
that arises from the dynamical interplay between the fluctuations in the optical mode and the mechanical motion.
If we write the self energy as $\Sigma(\omega \simeq \omega_j)= -i 2g^2_j \omega_j  \eta_c (\omega\simeq \omega_j)$,
for $j \equiv x,y$,   its imaginary part related to the optomechanical damping, while its  {\em real} part is related to
the optical spring shift squared:
\begin{eqnarray}
(\delta \omega_{(OS,x)})^2 &\simeq & \omega_x \text{Re}{ \left\lbrace -i 2g^2_x \eta_c (\omega=\omega_x)\right\rbrace } \nonumber \\
(\delta \omega_{(OS,y)})^2 &\simeq& \omega_y \text{Re}{\left\lbrace - i 2g^2_y \eta_c (\omega=\omega_y)\right\rbrace}
\label{OS}
\end{eqnarray}
For red detuning, it softens the
mechanical frequencies. It is strongest at the node and tends to zero as the antinode is approached.
In \cite{MTTM2020}, an additional correction was found, resulting from the intrinsic 2D dynamics, that here is negligible.

\section{The $S_{xy}$ spectrum}
It was shown in  \cite{MTTM2020} that one can correct the 1D mechanical displacement spectra
 to allow  for hybridisation. For the $x$ displacement, for example, one can write:
\begin{equation}
\hat{x}^\text{3D}(\omega) =  \hat{x}^\text{1D}(\omega) +\mathcal{R}_{xy}(\omega)\hat{y}^\text{3D}(\omega)+\mathcal{R}_{xz}(\omega)\hat{z}^\text{3D}(\omega) \label{3DNoise}
\end{equation}
using the appropriate hybridisation functions, $\mathcal{R}_{jk}(\omega)$, introduced in  \cite{MTTM2020}, to correct the unhybridised
spectra, $\hat{x}^\text{1D}(\omega)$.
Analogous expressions are obtained for $y,z$. Since the unhybridised spectra are well-known and given in terms of optical noises, the $x,y,z$ expressions can be rearranged and solved in closed form to obtain PSDs and arbitrary
correlation spectra. It is assumed that linearisation of
the equations of motion is valid and we can use  quantum linear theory (QLT). 

We obtained `exact' QLT PSDs $S_{xx} (\omega),S_{yy} (\omega)$
and correlation spectra $S_{xy} (\omega)$ to compare with experiment.
However, for physical insight, we also obtain below a simplified analysis  that  gives excellent 
agreement with the full 3D QLT. As the motion is approximately 2D, we
neglect the $z$ motion and  we further approximate:
\begin{alignat}{1}
\hat{x}^{\text{3D}}(\omega) & \simeq \hat{x}^\text{1D}(\omega)+\mathcal{R}_{xy}(\omega)\hat{y}^{\text{1D}}(\omega)\\
\hat{y}^{\text{3D}}(\omega) & \simeq \hat{y}^\text{1D}(\omega)+\mathcal{R}_{yx}(\omega)\hat{x}^{\text{1D}}(\omega).
\end{alignat}
where for a modest hybridisation correction, we substitute the 1D expressions  in the last term. 
 Below, we drop the 3D superscript and assume that $\hat{x},\hat{y} $ inlude hybridisation.
 To compare with experiment, we consider the symmetrised mechanical correlations:

 \begin{equation}
S_{xy} (\omega)= \frac{1}{2} \left  \langle [ \hat{x}]^\dagger \hat{y}\rangle + \langle [ \hat{y}]^\dagger \hat{x}\rangle \right)
\end{equation}
The current experiments are in weak coupling regimes and there is little mechanical-optical hybridisation.
This means that  $   \langle [ \hat{x}^{1D}]^\dagger \hat{y}^{1D} \rangle \simeq 0$. But in regimes of strong optical back-actions
even the unhybrised modes can be correlated hence $   \langle [ \hat{x}^{1D}]^\dagger \hat{y}^{1D} \rangle \neq 0$. We have verified numerically that the 1D components have negligible  cross correlations, so $ S_{xy}^\text{1D}\simeq 0$  and hence:
\begin{equation}
S_{xy} (\omega) \approx  \left( \text{Re}({\mathcal{R}_{yx}}(\omega)) S_{xx}(\omega)  +\text{Re}({\mathcal{R}_{xy}}(\omega)) S_{yy} (\omega)\right)
\end{equation}
 We can see that the cross-corrrelations are closely related to the real part of the hybridisation functions.
The latter can be given explicitly:

\begin{eqnarray}
\mathcal{R}_{xy}(\omega)&=& \frac{i\mu_{x}(\omega)}{M_{x}(\omega)} G(\omega)  \ \textrm{and} \nonumber \\
\mathcal{R}_{yx}(\omega)&=& \frac{i\mu_{y}(\omega)}{M_{y}(\omega)} G(\omega)
\end{eqnarray}

where $G(\omega)= \left[i\eta_{c}(\omega)g_{x}g_{y}+g_{xy}\right]$ is a term that represents the interference between the
`direct' static coupling  between $x$ and $y$ (proportional to $ g_{xy}$); and an indirect, cavity mediated, coupling term (
proportional to $g_{x} g_{y}$). The prefactors $M_{j}(\omega)=1+g_{j}^{2}\mu_{j}(\omega)\eta_c(\omega)$ include a small 
optical backaction correction to each displacement.
For our simplified analysis, we take $M_{j} \simeq 1$.  Numerical tests showed this is
an excellent approximation. The reason for this is that the small backaction correction 
 is peaked around each of the mechanical frequencies, i.e. at $M_x(\omega \approx \omega_x)$, and $M_y(\omega \approx \omega_y)$
 while for the cross-correlation, we show below the values around $M_x(\omega \approx \omega_y)\approx 1$, and $M_y(\omega \approx \omega_x) \approx 1$
are most important. 

 We note that, in the present discussion, we  refer to both the cavity mediated couplings $\eta_{c}(\omega)g_{x}g_{y}$
as well as the usual optomechanical back-action terms $g_j^2 \eta_0(\omega)$ as  `optical backaction' terms, but clearly, in the former case, the optical backaction acts on different mechanical modes.

{ \bf Optical and mechanical susceptibilities:} The $\mu_{j}(\omega)$ are mechanical susceptibilities, while $\eta_c$ is the optical susceptibility. We have the usual mechanical susceptibility $\mu_j(\omega)=\chi(\omega,\omega_j)-\chi^*(-\omega,\omega_j)$ and optical susceptibility  $\eta_c (\omega)=\chi(\omega,-\Delta)-\chi^*(-\omega,-\Delta)$, where  eg
$\chi(\omega,\omega_x)=[-i(\omega-\omega_x)+\frac{\Gamma}{2}]^{-1}$ and
 $\chi(\omega,\Delta)=[-i(\omega-\Delta)+\frac{\kappa}{2}]^{-1}$

\subsection{Anticorrelation of $S_{xy}$}

We can readily show that the $x-y$ modes are in general  anti-correlated (and this was observed in 
the experimental data)
by showing that the hybridisation functions $ \text{Re}{\mathcal{R}_{xy}} \approx -\text{Re}{\mathcal{R}_{yx}} $.

We have shown that  $\mathcal{R}_{xy}(\omega)\simeq G(\omega) i\mu_{x}(\omega)$ and $\mathcal{R}_{yx}(\omega) \simeq G(\omega) i\mu_{y}(\omega)$ and:
\begin{equation}
S_{xy} (\omega) \approx  \left( \text{Re}\left[ i\mu_y(\omega) G(\omega)  \right] S_{xx}    + \text{Re}\left[i\mu_x(\omega)G(\omega)  \right] S_{yy} \right).
\end{equation}

However, the PSDs are sharply peaked about the mechanical frequencies
$S_{xx}(\omega \approx \omega_x)$ and $S_{yy}(\omega \approx \omega_y)$,
hence we are interested in the value prefactors at those frequencies, namely
$\mu_y(\omega\sim \omega_x) G(\omega\approx \omega_x)$ and $\mu_x(\omega\sim \omega_y) G(\omega \approx \omega_y)$.

Since $\kappa \gg |\omega_x-\omega_y|$, the cavity susceptibility function $\eta_{c} (\omega)$, and hence
 $G(\omega)$ is insensitive to frequency: i.e. $G(\omega_x) \sim G(\omega_y)$.
Hence the anticorrelation behavior must originate in the mechanical susceptibilities.

However, since $\omega_x+\omega_y \gg |\omega_x-\omega_y|$ and the mechanical damping 
$\Gamma \ll |\omega_x-\omega_y|$ is negligible at ultrahigh vacuum,
we can write:
 \begin{equation}
\mu_{x} (\omega \sim \omega_y) \simeq [-i(\omega_y-\omega_x)+\frac{\Gamma}{2}]^{-1} \simeq i/(\omega_x-\omega_y)\\
\end{equation}
and
\begin{equation}
\mu_{y} (\omega  \sim \omega_x) \simeq [-i(\omega_x-\omega_y)+\frac{\Gamma}{2}]^{-1} \simeq i/(\omega_y-\omega_x)\\
\end{equation}

\begin{figure}[t]
{\includegraphics[width=3.3in]{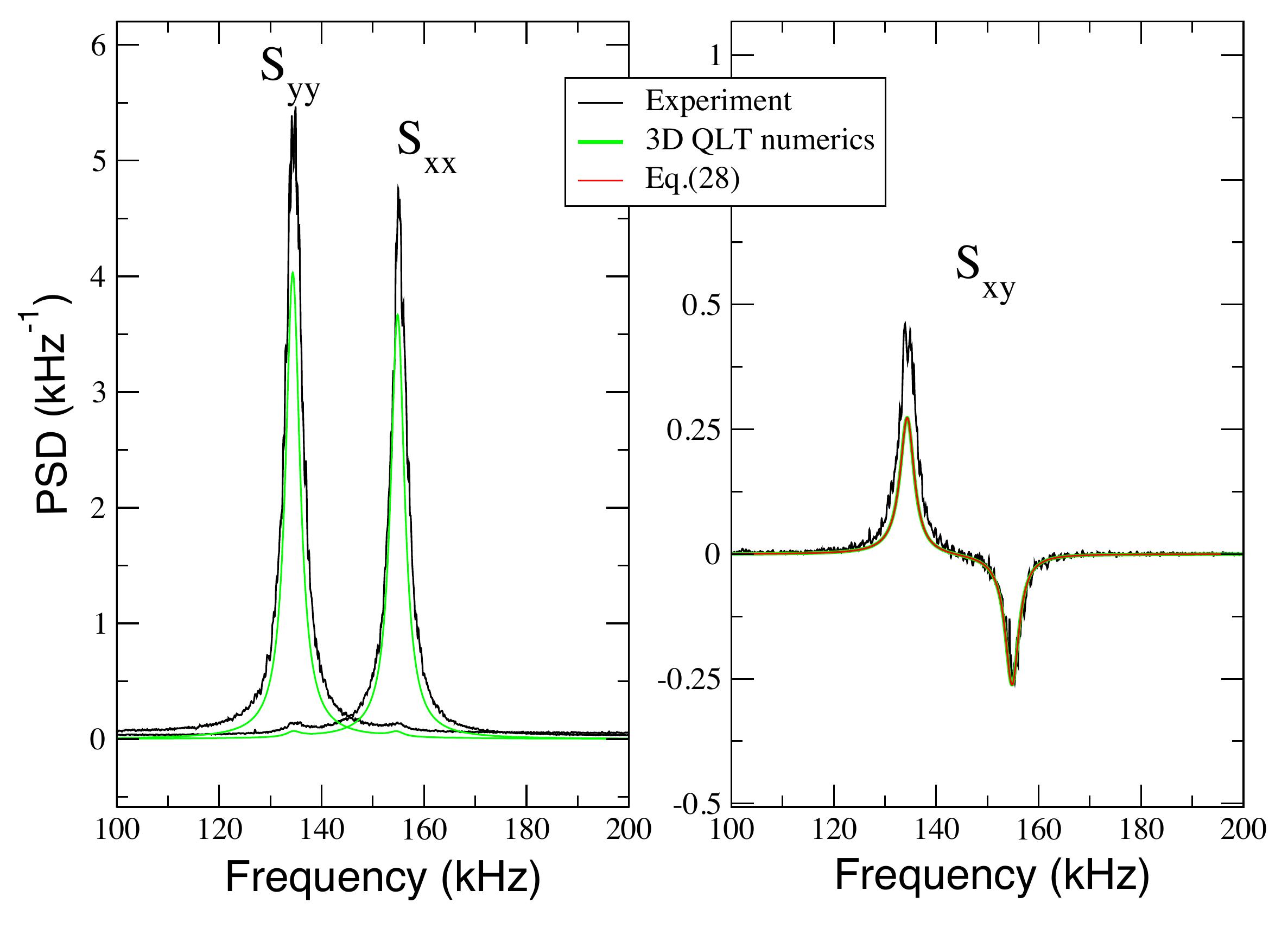}} \caption{  \textbf{(left panel)} PSDs for $x$ and $y$  for $\Delta/2\pi=-176$ kHz,
near the node, for experiment and QLT numerics. \textbf{(right panel)} for the same parameters, plots the cross correlation 
spectra, showing that the approximate expression Eq.\ref{Sxyapprox} gives excellent agreement with the full theory as well as the
experimental $x$ peak; the experimental $y$ peak requires a systematic correction for all data points as discussed in the main text. }
\label{Fig4}
\end{figure}

Finally,  we obtain an approximate expression for the mechanical correlation spectra:

\begin{equation}
S_{xy} (\omega) \approx  \frac{G(\omega)}{\omega_x-\omega_y} [  S_{yy}(\omega) - S_{xx}(\omega) ]\label{Sxyapprox}
\end{equation}
showing clearly that the $x$ and $y$ peaks have opposite signs. So the anticorrelation arises because the susceptibility for $y$ involves upconverting in frequency whereas $x$ represents a downconversion. 
The overall sign flips when $G(\omega)$ changes sign. For $\theta=3\pi/4$ then $g_x=-g_y$
and $g_{xy} \to -g_{xy}$  so there is a global sign flip of $S_{xy}$ relative to $\theta=3\pi/4$.

In the figure we test this simple expression against the exact QLT anticorrelation spectrum.
We can see that there is remarkable agreement everywhere: the expression is remarkably accurate. Since $G(\omega \sim \omega_x) \simeq G(\omega \sim \omega_y) \equiv G$, the prefactor is a constant and we can equate it to the rotation angle $\Phi \sim  \frac{G}{\omega_x-\omega_y}$. 
In practice, comparisons with experimental spectra employ the average
of the full QLT
expressions $\Phi=\frac{1}{2} \left( \text{Re}[{R}_{xy}] -  \text{Re}[{R}_{yx}] \right)$,
but  the simplified expressions are generally quite accurate.\\

\subsection{Suppression of hybridisation}

  If the term $G(\omega) \left[i\eta_{c}(\omega)g_{x}g_{y}+g_{xy}\right] \simeq 0$, the destructive interference between $x-y$ coupling and indirect, cavity-mode mediated coupling suppresses hybridisation and hence
$S_{xy} \simeq 0$.

Since the direct coupling $g_{xy}\simeq-g_{x}g_{y}\frac{2\text{Re}(\bar{\alpha})\cos{\phi}}{E_{d}\sin^{2}{\phi}}$, and $\bar{\alpha}\simeq -iE_{d}\text{cos}(\phi)[i\Delta-\kappa/2]^{-1}$:
\begin{equation}
g_{xy}\simeq g_{x}g_{y}\left[\frac{2\Delta\cot^{2}{\phi}}{\Delta^{2}+\frac{\kappa^{2}}{4}}\right].\label{gjk}
\end{equation}
Thus depending on the positioning, $\Delta$ or $\kappa$, the direct
couplings contribution can be similar or exceed the cavity mediated
coupling. Direct and indirect contributions, in general,interfere
destructively. We can show that $i\eta_c(\omega)\to\frac{-2\Delta}{(\kappa/2)^{2}+\Delta^{2}}$
if $\text{-}\Delta\gg\omega$ (and we are interested primarily in
the region $\omega\sim\omega_{j}$). Thus for large $-\Delta$:

\begin{equation}
G(\omega)\approx g_{x}g_{y}\left[\frac{-2\Delta}{\Delta^{2}+(\kappa/2)^{2}}\right]\left[1-\cot^{2}\phi\right],\label{xycouple}
\end{equation}
and we see that the $G$ is real and frequency independent. Furthermore, at $\phi=\pi/4$  the $x$-$y$ hybridisation almost fully vanishes so we have a cancellation point where the $S_{xy}$ correlation
spectra are near zero.

\begin{figure}[ht!]
    {\includegraphics[width=3in]{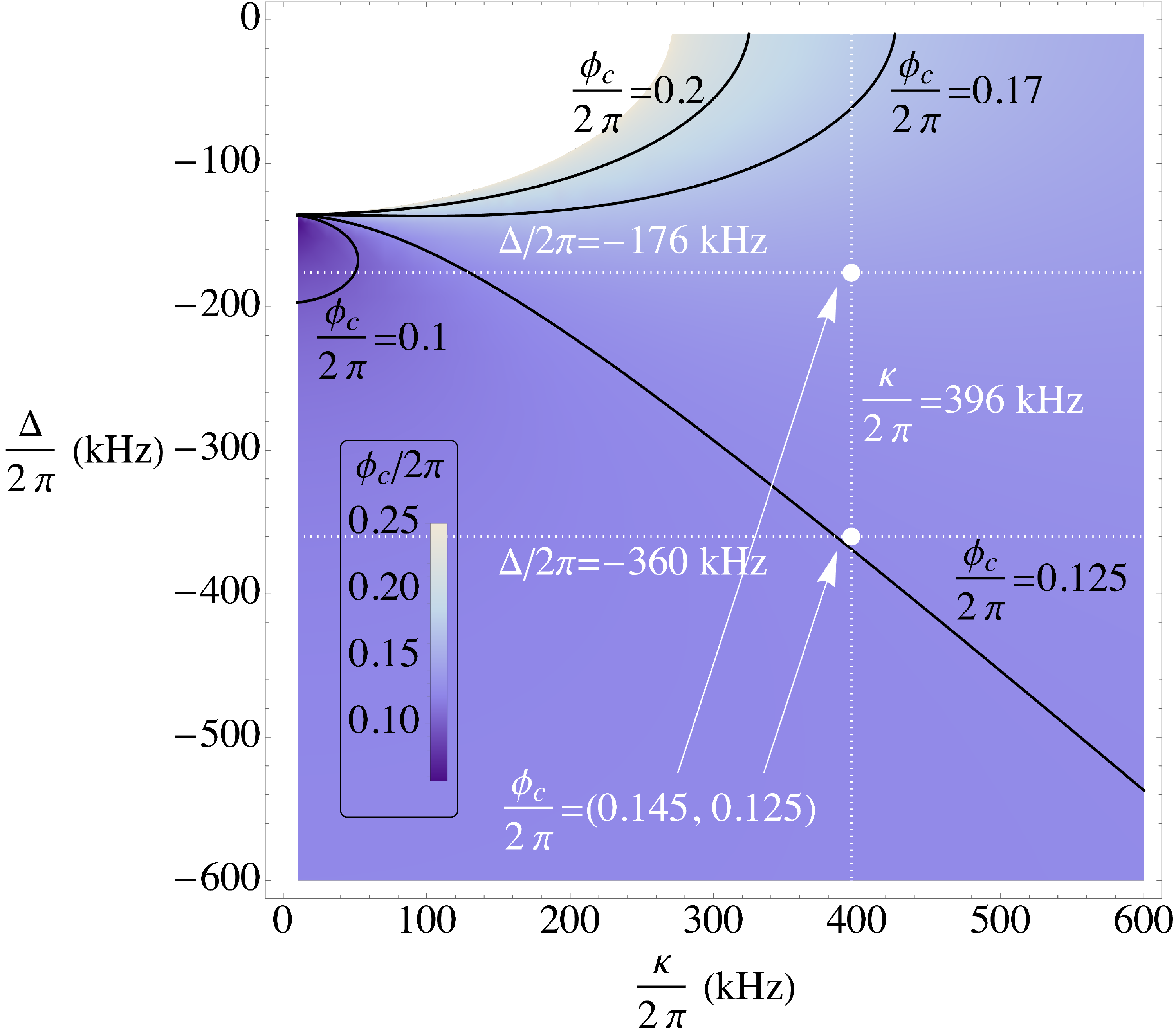}} \caption{
    Density plot of the cancellation point $\phi_c = \tan^{-1}(\sqrt{C_\phi})$ as a function of the cavity linewidth $\kappa$ and optical detuning $\Delta$, with black contours at notable values.
    $C_\phi$ is shown in Eq.\ref{eq:C_phi}.
    White-dashed lines show the experimental parameters, with white dots representing the two sets of data taken where $\phi_c$ was investigated.
    As highlighted by Fig.2 in the main text, the $\phi_c$ predictions agree well with experimental observations.
    The blank ellipse in the top left is the case where $C_\phi < 0$, leading to complex-valued $\phi_c$ that is not plotted.
    This condition leads to $\Delta^2 + \kappa^2/4 < \omega_y^2$, showing the ellipse's radii to be $2\omega_y$ and $\omega_y$ in the $\kappa$ and $\Delta$ axes, respectively.
    The large detuning limit of $\phi_c/2\pi \to 0.125$ can be seen in the lower half.
    The plot suggests that experiments with this cavity ($\kappa = 396\,\text{kHz} \cdot 2\pi \simeq 2.9~\omega_y$) are limited to $\phi_c/2\pi \sim 0.17$, even in a very small detuning limit $-\Delta \ll \omega_y$.
    Nevertheless, the higher-valued (brighter) area around the $\phi_c/2\pi = 0.2$ contour suggests that a better cavity with a realistically lower $\kappa \gtrsim 2\omega_y$, may have cancellation closest to the node with $\phi_c/2\pi \sim 0.2$, so long as the detuning is below resonance. However, if the detuning is above resonance, the $\phi_c/2\pi = 0.1$ contour highlights a region where an increasingly perfect cavity $\kappa \ll \omega_y$ may even move $\phi_c$ toward the antinode.
}
\label{Fig5}
\end{figure}

\section{Behavior  of $\phi=\phi_c$}

In the present experimental study we are not necessarily in the large detuning limit
so we consider other regimes including  $-\Delta \sim \omega_{x,y}$ and find
 the cancellation $\phi=\phi_c$ moves towards  the node.

Although the present experimental data is not conclusive, it is certainly consistent with a
$\phi_c/2\pi \simeq 0.15$  for $-\Delta/2\pi=176$ kHz, closer to the node than for $-\Delta/2\pi=360$ kHz,
for which $\phi_c/2\pi \simeq 0.125$.  Operating even closer to the node would be even more  advantageous for quantum optomechanics as it reduces cavity photon occupancies 
and thus the deleterious effects of optical noise.

In the present study, we consider only mode rotation in the $x-y$ plane. For experiments in the good cavity limit and nearer resonance, a model of mode rotation must consider the tripartite interaction between  $x$, $y$ and the optical mode.
Below we neglect optical mixing and present a simplified analysis where the mode rotation angle does not consider 
the neglected $   \langle [ \hat{x}^{1D}]^\dagger \hat{y}^{1D} \rangle \simeq 0$ terms.  

We analyse simply the behavior of of the simplified model $\Phi(\phi)  \sim  \frac{G}{\omega_x-\omega_y}$. We write:

\begin{equation}
\Phi(\phi)
=
\frac{A}{\omega_y - \omega_x}
\left(\text{Re}[i\eta_\text{c}(\omega_y)]\sin^2\phi + B\cos^2\phi\right)
\label{eq:Phi_phi}
\end{equation}
where
$A
=
\frac{g_x g_y}{\sin^2\phi}
=
E_\text{d}^2 k^2 X_\text{zpf} Y_\text{zpf} \sin\theta \cos\theta $,
and
$B
=
\frac{g_{xy}}{g_x g_y}
=
\frac{2\Delta}{\Delta^2 + \kappa^2/4}$.

Solving for $\Phi(\phi_c) = 0$ leads to:

\begin{equation}
\phi_c = \tan^{-1} \left( \sqrt{C_\phi} \right)
\label{eq:phi_c}
\end{equation}
where $C_\phi = -B/\text{Re}[i\eta_\text{c}(\omega_y)]$.

\subsection{Moving $\phi_c$ closer to the node}

We plot Eq.\ref{eq:phi_c} in {\bf Fig.\ref{Fig5}}, showing how the cancellation point $\phi_c$ varies with $\kappa$ and $\Delta$.
First, we highlight that the predictions for the two experimentally explored sets of parameters (white dots) match the observation of the cancellation point moving closer to node as the detuning approaches resonance $-\Delta/2\pi = 176\,\text{kHz} \sim \omega_y/2\pi = 136\,\text{kHz}$.

Next, we note that {\bf Fig.\ref{Fig5}} (and hence Eq.\ref{eq:phi_c}) verifies that, as $C_\phi$ gets larger ($\kappa,\Delta \gg \omega_y$), $\phi_c/2\pi \to 0.125$, consistent with the analysis from the previous section.

Eq.\ref{eq:phi_c} also shows that $\phi_c$ is real-valued as long as $C_\phi \geq 0$.
Using the full form of the cavity susceptibility function $\eta_\text{c}(\omega)$, we obtain:

\begin{equation}
C_\phi
=
\frac
{(\Delta^2 + \kappa^2/4 - \omega_y^2)^2 + (\kappa\omega_y)^2}
{(\Delta^2 + \kappa^2/4 - \omega_y^2)(\Delta^2 + \kappa^2/4)}
\label{eq:C_phi}
\end{equation}
By inspection, it is seen from the first parentheses in the denominator (other terms being always positive) that $C_\phi \geq 0$ holds so long as $\Delta^2 + \kappa^2/4 > \omega_y^2$.
This condition is seen clearly in {\bf Fig.\ref{Fig5}} to be the form of the blank ellipse in the top left, which has a radius in the $\kappa~(\Delta)$ axis of $2\omega_y~(\omega_y)$.

Lastly, the higher-valued area between the $\phi_c/2\pi = 0.2$ and $0.17$ contours suggests that --  although this particular $\kappa/2\pi = 396\,\text{kHz}$ might be limited to $\phi_c/2\pi \sim 0.17-0.18$ (even in an extreme $-\Delta \ll \omega_y$ limit) -- for operation in the good cavity regime, with  $\kappa \gtrsim 2\omega_y \simeq 270\,\text{kHz}\cdot2\pi$, then $\phi_c/2\pi \sim 0.2$ may be possible with a detuning below resonance.

A future analysis will investigate how best to extract the mode orientation in the $x-y$ plane in the good cavity regime and
with strong light-matter coupling.

\end{document}